\documentclass[aps,showpacs]{revtex4}
\usepackage{latexsym,amsmath,amssymb,epsfig}
\newcommand{\bea}{\begin{eqnarray}}
\newcommand{\eea}{\end{eqnarray}}
\newcommand{\be}{\begin{equation}}
\newcommand{\ee}{\end{equation}}
\newcommand{\beast}{\begin{eqnarray*}}
\newcommand{\eeast}{\end{eqnarray*}}
\newcommand{\pkt}{\; .}
\newcommand{\kma}{\; ,}

\newcommand{\labelcaption}[2]{\caption[#1]{\label{#1}#2}}

\newcounter{subequation}[equation]
\makeatletter
\expandafter\let\expandafter\reset@font\csname reset@font\endcsname
  {\endeqnarray\stepcounter{equation}}
\makeatother

\begin{document}
\rightline{gr-qc/0602039}
\date{\today}
\title{The number of negative modes of the oscillating bounces}
\author{George Lavrelashvili}
\email{lavrela@itp.unibe.ch}
\affiliation{Department of Theoretical Physics,
University of Geneva,
24 quai Ernest-Ansermet,
CH 1211 Geneva 4,
Switzerland\\
and\\
Department of Theoretical Physics,
A.Razmadze Mathematical Institute,
GE 0193 Tbilisi, Georgia}

\begin{abstract}
\noindent
The spectrum of small perturbations about oscillating bounce solutions
recently discussed in the literature is investigated.
Our study supports quite intuitive and expected result:
the bounce with $N$ nodes has exactly $N$ homogeneous negative modes.
Existence of more than one negative modes makes obscure the relation
of these oscillating bounce solutions to the false vacuum decay processes.

\end{abstract}
\pacs{98.80.Jk, 98.80.Cq, 04.62.+v}
\maketitle

\section{Introduction}
\par

Having in mind recent progress in string theory, predicting string
landscape (many vacua) picture, it is very important to have deeper
understanding of metastable vacuum decay processes with gravity taken into account.
Metastable (false) vacuum decay is a quantum tunnelling process and
is usually described within the Euclidean approach \cite{col77}.
It was shown that the metastable vacuum decay in flat space-time proceeds via true vacuum bubbles
nucleation in the false vacuum and subsequent growth of these bubbles.
Bubble nucleation process is described by bounce, classical solution of the Euclidean equations of motion
with certain boundary conditions. It was found \cite{coglma78} that in flat space-time
the $O(4)-$ symmetric bounce has lowest action and gives main contribution to the tunnelling.
Furthermore, it was shown that there is exactly one negative mode in spectrum of small
perturbations about bounce solution in flat space-time \cite{caco77}.
This negative mode is very essential and makes decay picture coherent.
In Coleman's words: ``There may exist solutions in other ways like bounces and
which have more than one negative eigenvalue, but, even if they do exist,
they have nothing to do with tunnelling" \cite{col88}.

Bounce solution in presence of gravity was found by Coleman and De Luccia \cite{colu80}.
In addition some exited multi bounce solutions are known in the literature:
Bousso and Linde discussed double-bubble instantons \cite{bl98} and more recently
the oscillating bounce solutions were studied in details
by Hackworth and Weinberg \cite{hw05,wein05}. It was found that multi bounces have higher
Euclidean action than the bounce itself. So, it was suggested that the multi bounces give
sub-leading contribution to the vacuum decay processes.

The aim of the present letter is to investigate the number of negative modes
of these oscillating bounce solutions
in order to check their relevance to the tunnelling.
While in flat space-time finding a negative mode about bounce is straightforward task,
when gravity is taken into account it is more involved problem
\cite{lrt85,tasa92,lav98,tanaka99,klt00,lav00,grtu01}.
It was shown that with the proper reduction procedure one finds a single
negative mode about Coleman-De Luccia bounce \cite{klt00}.

The rest of the paper is organized as follows: in the next section we discuss the Euclidean
equations of motion and boundary conditions for the bounce solution. In Section III
we present Schr\"oedinger equation for linear perturbations about the bounce and
in the Section IV be show out numerical results for concrete choice of scalar
field potentials.

\section{Bounce solution} \label{bounce}
\par

Let's consider the theory of a scalar field coupled to gravity
which is defined by the following Euclidian action
\be
S_E=\int {d^4x\sqrt{g} \; \Bigl(\frac{1}{2}\nabla_\mu\phi \nabla^\mu\phi
+ V(\phi)-\frac{1}{2\kappa}R \Bigr)} \kma
\ee
where $\kappa=8\pi G_{N}$ is the reduced Newton's gravitational constant.

The most general $O(4)$ invariant metric is parameterized as
\be\label{metric}
ds^2=N^2(\sigma)d\sigma^2+a^2(\sigma)d\Omega_3^2 \kma
\ee
where $N(\sigma)$ is the Lapse function, $a(\sigma)$ is the scale factor
and $d\Omega_3^2$ is metric of unit three-sphere:
\be
d\Omega_3^2 = d\chi^2 + {\rm sin}^2\chi (d\theta^2
+{\rm sin}^2(\theta) d\varphi^2) \pkt
\ee
For metric Eq.~(\ref{metric}) the curvature scalar looks like
\be
R= \frac{6}{a^2}-\frac{6 \dot{a}^2}{a^2 N^2}-\frac{6 \ddot{a}}{a N^2}
+ \frac{6 \dot{a} \dot{N}}{a N^3} \kma
\ee
where $\dot{} = d/ d\sigma$.
Using ansatz Eq.~(\ref{metric}) and assuming that $\phi=\phi(\sigma)$
we get the reduced action in the form
\be
S_E=S_E(\phi,N,a)= 2\pi^2 \int d\sigma \Bigl(
\frac{a^3}{2N}\dot{\phi}^2 + a^3 N V(\phi)
-\frac{3 a N}{\kappa} +\frac{3 a \dot{a}^2}{\kappa N}
+\frac{3 a^2 \ddot{a}}{\kappa N}
-\frac{3 a^2 \dot{a}\dot{N}}{\kappa N^2} \Bigr) \pkt
\ee
Corresponding field equations in the proper time gauge, $N=1$, are
\be
\ddot{\phi}+3\frac{\dot{a}}{a}\dot{\phi}=\frac{\partial V}{\partial\phi} \kma
\ee
\be
\ddot{a}=-\frac{\kappa a}{3} (\dot{\phi}^2+V(\phi)) \kma
\ee
\be
\dot{a}^2= 1+ \frac{\kappa a^2}{3}(\frac{\dot{\phi}^2}{2}-V) \pkt
\ee
Now let's assume that potential $V(\phi)$ has two non-degenerate local minima
at $\phi=\phi_{\rm tr}$ and $\phi=\phi_{\rm fv}$, with $V(\phi_{\rm fv})>V(\phi_{\rm tr})$,
and local maximum for some $\phi=\phi_{\rm top}$, with $\phi_{\rm fv}<\phi_{\rm top}<\phi_{\rm tv}$,
Fig.~1. Euclidean solution describing vacuum decay - bounce - satisfies
these equations and in case $V(\phi) > 0$ has following boundary conditions
\be
\phi (0)= \phi_0,\qquad \dot{\phi}(0) = 0,\qquad a(0)=0, \qquad \dot{a}(0)=1 \kma
\ee
at $\sigma=0$ and
\be
\phi (\sigma_{max})= \phi_{m},\qquad \dot{\phi}(\sigma_{max}) = 0,\qquad
a(\sigma_{max})=0,\qquad \dot{a}(\sigma_{max})=1 \kma
\ee
at some $\sigma=\sigma_{max}$.
This assumes the following Taylor series at $\sigma \to 0$
\be
\phi(\sigma)=\phi_0+\frac{1}{8}{\frac{\partial V}{\partial\phi}}|_{\phi=\phi_0}\sigma^2
+\frac{1}{192}\frac{\partial V }{\partial\phi}|_{\phi=\phi_0}
(\frac{\partial^2V}{\partial \phi^2}|_{\phi=\phi_0}
+\frac{2\kappa V(\phi_0)}{3})\sigma^4 + O(\sigma^6) \kma
\ee
\be
a(\sigma)=\sigma-\frac{\kappa}{18}V(\phi_0)\sigma^3
-\frac{\kappa}{120}(\frac{3}{8}(\frac{\partial V}
{\partial\phi}|_{\phi=\phi_0})^2
-\frac{\kappa}{9}V^2(\phi_0))\sigma^5+O(\sigma^7) \kma
\ee
and similar power law behavior for non-singular bounces for $x\to 0$,
where $x=\sigma_{max}-\sigma$.

Whereas bounce solution always exists in the flat space-time,
when gravity is switched on, the existence of bounce depends on details of
scalar field potential. For wide class of potentials the existence of
bounce solution is determined by the value of parameter $\beta$ \cite{js89}
\footnote{For very flat potentials one needs more detailed investigation \cite{hw05}.},
\be
\beta=|V''(\phi_{\rm top})| / H^2 \kma
\ee
where $H^2={\kappa V(\phi_{\rm top})}/{3}$.
For $\beta < 4$ no Coleman-De Luccia bounce exists in the given potential.
Increasing $\beta$ more and more oscillating bounce solutions appear.
For broad class of potentials for a given $\beta$ there are oscillating bounces with up
$N$ nodes, where $N$ is the largest integer such that $N (N+3)< \beta$ \cite{hw05}.
In addition one finds also the Hawking-Moss solution \cite{hm82}
which exist in any potential with positive local maximum.

\section{Linear perturbations} \label{perturbations}
\par

The investigation of perturbations about the bounce solution is convenient to perform
in conformal frame \cite{klt00,lav00}.
Let's expand the metric and the scalar field
over a $O(4)-$ symmetric background as follows
\be
ds^2= a(\tau)^2 \Bigl((1+2 A(\tau))d\tau^2 +(1-2 \Psi(\tau))d\Omega_3^2 \Bigr)
\kma \qquad \phi=\varphi(\tau) + \Phi(\tau) \kma
\ee
where $\tau$ is the conformal time, $a$ and $\varphi$ are the background field values and
$A, \Psi$ and $\Phi$ are small perturbations.
In what follows we will be interested in the lowest (only $\tau$ dependant, `homogeneous') modes
and consider only scalar metric perturbations,
while the negative energy states are found previously exactly in this sector.

Expanding the total action, keeping terms up to the second order in perturbations
and using the background equations of motion we find
\be
S= S^{(0)}[a,\varphi]+S^{(2)}[A,\Psi,\Phi] \kma
\ee
where $S^{(0)}$ is the action of the background solution and $S^{(2)}[A,\Psi,\Phi]$
is the quadratic action. The Lagrangian corresponding to this quadratic action
is degenerate and describes constrained dynamical system.
Applying Dirac's formulation of generalized Hamiltonian dynamics
we get unconstrained quadratic action in the form \cite{klt00,lav00}
\be \label{conformal_qa}
S^{(2)}_E = 2\pi^2 \int \Bigl(\frac{1}{2}q'^2
+ \frac{1}{2}W[a(\tau),\varphi(\tau)]\; q^2 \Bigr) d\tau \kma
\ee
with the potential $W$ whose conformal time dependance is determined by the bounce solution
\cite{lav00}
\be
W[a(\tau),\varphi(\tau)]= \frac{a^2}{\cal Q} \frac{\delta^2 V}{\delta\varphi\delta\varphi}
-\frac{10 {a'}^2}{a^2 \cal Q}+\frac{12 {a'}^2}{a^2 {\cal Q}^2}
+\frac{8}{\cal Q}-6-3{\cal Q}
+\frac{\kappa a^4}{2 {\cal Q}^2}(\frac{\delta V}{\delta\varphi})^2
-\frac{2\kappa a a' \varphi'}{{\cal Q}^2} \frac{\delta V}{\delta\varphi}
\pkt
\ee
Here $q=a/\sqrt{\cal Q} \; \Phi$, prime denotes the derivative
with respect to conformal time $\tau$ and ${\cal Q}=1-\kappa \varphi'^2/6$.

Introducing new variable $f=\sqrt{a} q$ and passing to the proper time $\sigma$
quadratic action Eq.~(\ref{conformal_qa}) can be written in the form
\be \label{proper_qa}
S^{(2)}_E = 2\pi^2 \int \Bigl(\frac{1}{2}\dot{f}^2+ \frac{1}{2}U[a(\sigma),
\varphi(\sigma)] \; f^2 \Bigr) d\sigma \kma
\ee
with the potential $U$
\be
U[a(\sigma),\varphi(\sigma)]= \frac{1}{ Q} \frac{\delta^2 V}{\delta\varphi\delta\varphi}
-\frac{10 {\dot a}^2}{a^2 Q}+\frac{12 {\dot a}^2}{a^2 Q^2} +\frac{8}{a^2 Q}
-\frac{6}{a^2}-\frac{3Q}{a^2}-\frac{{\dot a}^2}{4 a^2}
+\frac{\kappa a^2}{2Q^2}(\frac{\delta V}{\delta\varphi})^2
-\frac{2\kappa a {\dot a}{\dot \varphi}}{Q^2} \frac{\delta V}{\delta\varphi}
-\frac{\kappa}{6} (\dot{\varphi}^2+V),
\ee
where $Q=1-\kappa a^2 {\dot \varphi}^2/6$.
So, spectrum of small perturbations about bounce solution is determined by the following
Schr\"odinger equation
\be \label{schroed_eq}
-\frac{d^2}{d\sigma^2} f + U[a(\sigma),\varphi(\sigma)] f = E f,
\ee
and the number of negative modes of the bounce solution is the number of bound states of these
Schr\"odinger equation.

\section{Numerical results} \label{numerics}
\par

Let's parameterize the general quartic scalar field  potential as follows:
\be
V=V_0+H^2 (-\frac{\beta}{2}\varphi^2-\frac{g}{3}\varphi^3+\frac{\lambda}{4}\varphi^4) \kma
\ee
with $H^2=\kappa V_0 /3$.

Passing to the dimensionless variables
\be
\tilde{\varphi}=\frac{\varphi}{v} \kma \tilde{a}=a v \kma \tilde{\sigma} =\sigma v \kma
\tilde{V}_0= \frac{V_0}{v^4} \kma \tilde{H}^2 =\frac{H^2}{v^2} \kma
\tilde{\kappa}= \kappa v^2 \kma
\ee
with $v^2={2\beta}/{\lambda}$ we will get the dimensionless equations of motion with the
rescaled potential (comp. \cite{hw05})
\be \label{V_pot_rescaled}
\tilde{V}=\tilde{V}_0+\tilde{H}^2 \beta (-\frac{1}{2}\tilde{\varphi}^2
- \frac{\tilde{g}}{3}\tilde{\varphi}^3+\frac{1}{2} \tilde{\varphi}^4)  \kma
\ee
where $\tilde{g}=g v/\beta$. In what follows we will use dimensionless variables and omit tildes.

Potential $U$ close to the $\sigma = 0$ behaves as
\be
U= \frac{3}{4 \sigma^2}+ U_0 + O(\sigma^2) \kma
\ee
where constant $U_0$ depends on the initial value of scalar field and
parameters of the background solution potential $V$.
For the potential Eq.~(\ref{V_pot_rescaled}) it is
\be
U_0=(-H^2 \beta-\frac{2}{3} \kappa V_0
-2 H^2 \beta g \phi_0
+6 H^2 \beta(1+\frac{\kappa}{18} ) \phi_0^2
+\frac{2}{9} \kappa H^2 \beta g \phi_0^3
-\frac{1}{3}\kappa H^2 \beta \phi_0^4) \pkt
\ee
The regular branch of the wave function $f$ behaves as
\be
f=\sigma^{3/2}(1+ \frac{1}{8}(U_0-E) \sigma^2 + O(\sigma^4)) \pkt
\ee

Convenient way to determine the number of bound states of Schr\"odinger equation
in a given potential is the investigation of the zero energy wave function.
The number of nodes of zero energy wave function exactly counts the number of negative
energy states \cite{aq95}.

Let's describe our results in details on concrete example.
For the parameters choice
\be
\kappa=0.001 \kma V_0=0.1 \kma \beta=70.03 \kma g=\frac{1}{2\sqrt{2}} \kma
\ee
the potential Eq.(\ref{V_pot_rescaled}) has local maximum at $\varphi_{\rm top}=0$,
metastable minimum at $\varphi_{\rm fv}=-0.6242212930$ and true vacuum at
$\varphi_{\rm tv}=0.8009979884$, Fig.~1.
There exists Coleman-De Luccia bounce solution in this potential,
oscillating bounces on top of it with up to $N=7$ nodes and, as always,
the Hawking-Moss solution \cite{hm82} with $\varphi \equiv \varphi_{\rm top}$.
Numerical investigation supports quite intuitive and expected result \cite{grtu01}:
the bounce with $N$ nodes has exactly $N$ negative modes.
Typical results are demonstrated for $N=3$ case on Fig.2 and Fig.3.
The zero energy wave function of Schr\"oedinger equation Eq.~(\ref{schroed_eq})
has in this case three nodes, which means that there are exactly three
negative energy states for $N=3$ oscillating bounce.
We also found this states explicitly and determined their energies:
$E_0=-0.0013787$, $E_1=-0.0004362$ and $E_2=-0.0001207$.
Corresponding Hawking-Moss solution has eight homogeneous negative modes,
which is consistent with chosen value of $\beta$.

To conclude, we found that the oscillating bounces studied in \cite{bl98} and \cite{hw05}
have more than one negative modes,
which makes their relevance to the tunnelling processes obscure.
So, question about contribution of oscillating bounces to the false vacuum decay amplitude
needs further investigation.

\section*{Acknowledgements}
\par
The main part of this work has been done during my visit to Geneva University, Switzerland
and paper was completed during short visit to the Albert-Einstein-Institute, Golm, Germany.
I would like to thank the theory groups of these institutions and especially Ruth Durrer
and Hermann Nicolai for kind hospitality.
It is a pleasure to thank the Tomalla foundation for the financial support.




\begin{figure}[b]
\centerline{
\epsfig{file=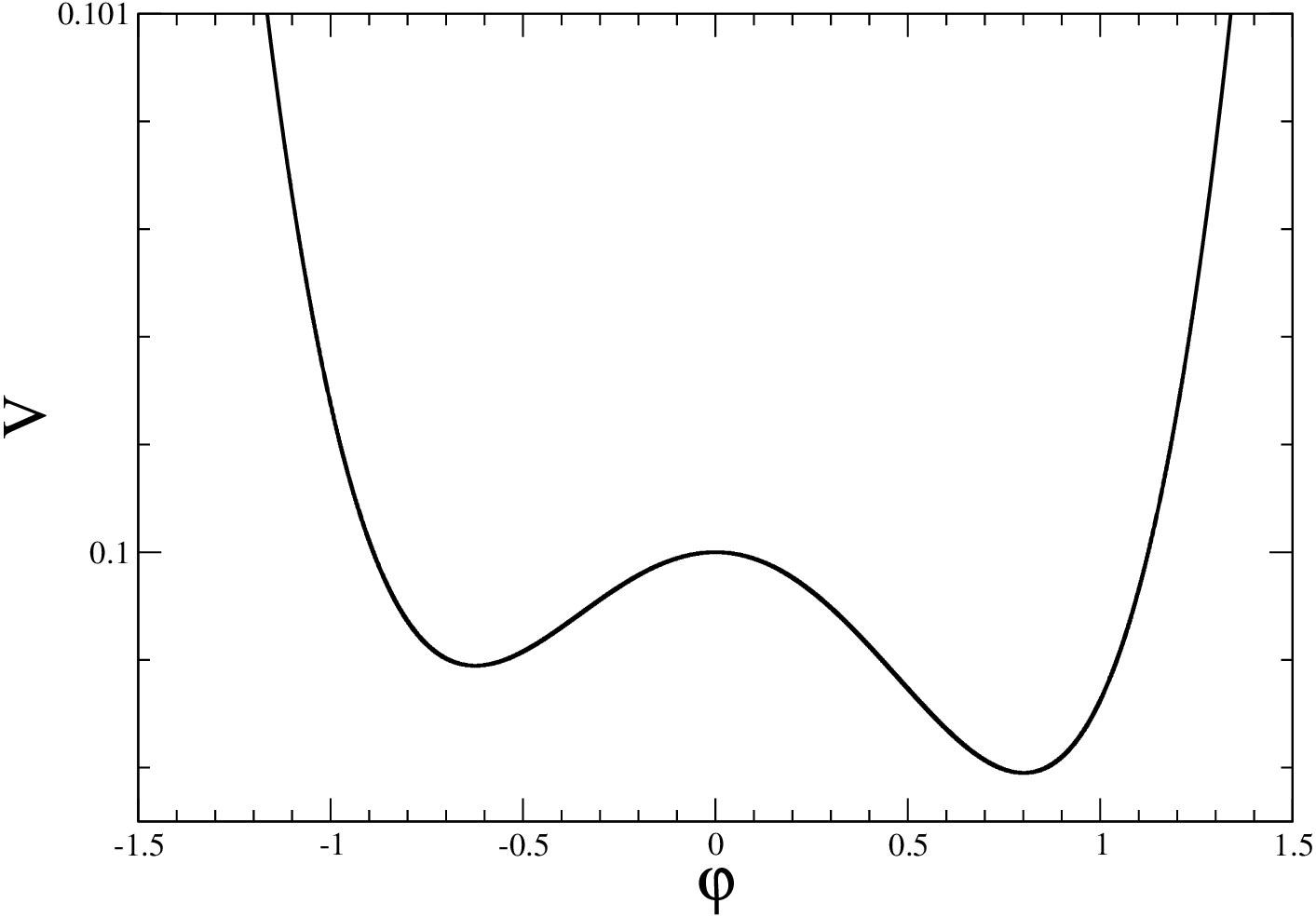,width=0.6\hsize,
bbllx=4.5cm,bblly=1.5cm,bburx=24.5cm,bbury=20.0cm}}
\labelcaption{pot}
{Scalar field potential $V(\varphi)$.}
\end{figure}

\begin{figure}
\centerline{
\epsfig{file=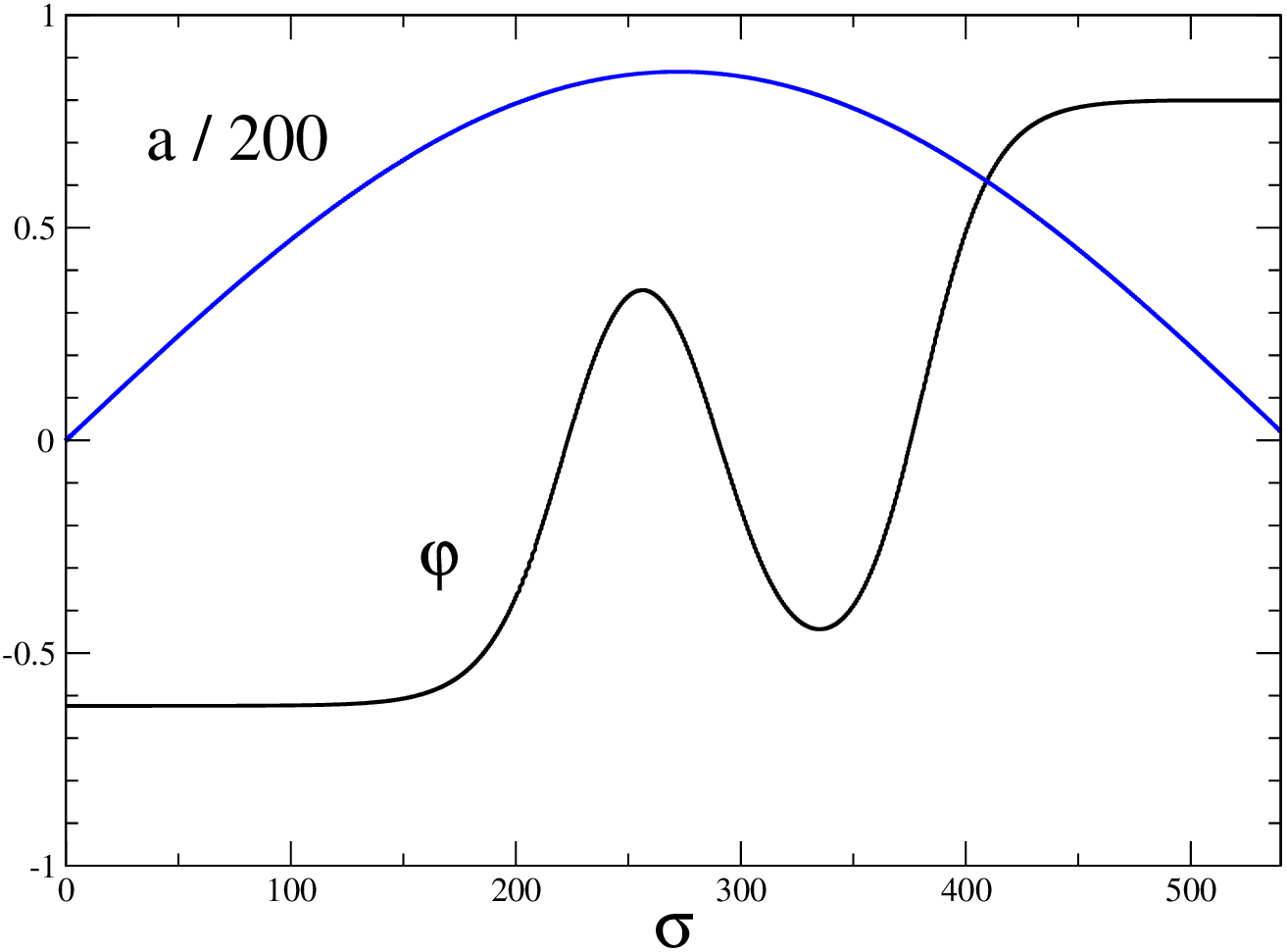,width=0.6\hsize,
bbllx=4.5cm,bblly=1.5cm,bburx=24.5cm,bbury=20.0cm}}
\labelcaption{a_phi}
{Oscillating bounce solution with three nodes of $\varphi$.}
\end{figure}

\begin{figure}
\centerline{
\epsfig{file=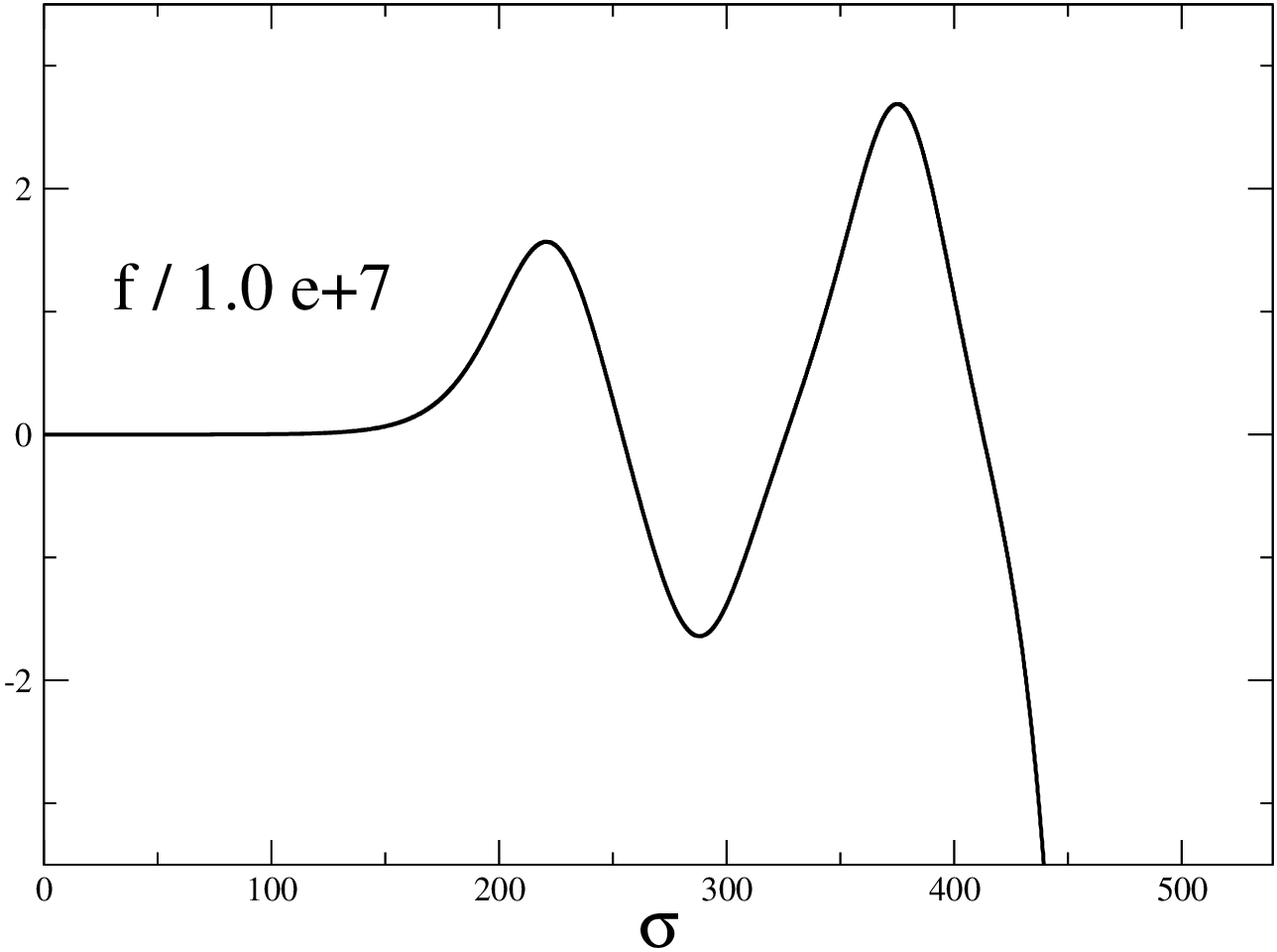,width=0.6\hsize,
bbllx=4.5cm,bblly=1.5cm,bburx=24.5cm,bbury=20.0cm}}
\labelcaption{q}
{Zero energy wave function $f$ of Schr\"odinger equation of linear
perturbations about oscillating bounce solution with three nodes of $\varphi$.}
\end{figure}

\end{document}